# GBM Returns the Best Prediction Performance among Regression Approaches: A Case Study of Stack Overflow Code Quality

Sherlock A. Licorish[1], Brendon Woodford[1], Lakmal Kiyaduwa Vithanage[1], Osayande Pascal Omondiagbe[2]
[1] University of Otago, Dunedin, Otago, New Zealand
[2] Australian Institute of Marine Science, PMB 3, Townsville, Queensland, Australia

*Abstract*—Practitioners are increasingly dependent on publicly available resources for supporting their knowledge needs during software development. This has thus caused a spotlight to be paced on these resources, where researchers have reported mixed outcomes around the quality of these resources. Stack Overflow, in particular, has been studied extensively, with evidence showing that code resources on this platform can be of poor quality at times. Limited research has explored the variables or factors that predict code quality on Stack Overflow, but instead has focused on ranking content, identifying defects and predicting future content. In many instances approaches used for prediction are not evaluated to identify the best techniques. Contextualizing the Stack Overflow code quality problem as regression-based, we examined the variables that predict Stack Overflow (Java) code quality, and the regression approach that provides the best predictive power. Six approaches were considered in our evaluation, where Gradient Boosting Machine (GBM) stood out. In addition, longer Stack Overflow code tended to have more code violations, questions that were scored higher also attracted more views and the more answers that are added to questions on Stack Overflow the more errors were typically observed in the code that was provided. Outcomes here point to the value of the GBM ensemble learning mechanism, and the need for the practitioner community to be prudent when contributing and reusing Stack Overflow Java coding resource.

## I. INTRODUCTION

Practitioners use platforms such as Stack Overflow and public software code repositories to support their software development efforts [1, 2]. While these platforms provide utility for the software development community, studies have shown that they can present content with questionable quality at times, which may lead to immediate software errors and long term defects [3, 4], [5]. Thus, a growing body of research is dedicated to studying these platforms [2] and providing mechanisms for detecting errors and improving the quality of content that is available to practitioners [6]. Researchers have also developed various datasets to support research efforts [7], where the academic and practitioner communities are able to experiment with novel modelling techniques [8].

Accordingly, research efforts have been dedicated to investigating various approaches to predict aspects of practitioners' processes [8], ranking the content on online platforms [9], identifying particular defects online [10] and predicting future content [11]. Various degrees of precision, recall and F1-score have been reported by these works, pointing to the potential utility of the modelling approaches that are explored.

Less effort is committed to comparing the actual techniques that are used however [12]. In fact, in many cases approaches are presented as fit for purpose, albeit only one dataset and one method is used for experimentation [13]. While the outcomes presented by these studies help us to understand the usefulness of the approaches that are used and contribute to our understanding of the state of the art, there is need to properly evaluate the methods that are available for prediction. In the context of Stack Overflow code quality, we know that the quality of content on this portal varies [1], but there is little insights available into the specific variables that influence the errors that are observed. In addition, while the community is keenly interested in such insights [14], it would be equally beneficial to learn about the most accurate methods for predicting this quality outcome. This is a regression-based problem, where little is known about how the simplest or most complex approaches perform [15], and when these approaches provide the most value for the computational resource that they use.

We exploit this opportunity in this study in bridging the gap. Instead of creating an approach for predicting Stack Overflow code quality, we first explore the variables that predict code quality before investigating the regression approach that provides the best predictive power. In the process we also examine those approaches that reported poorer performance. We contribute to the community's understanding of the quality of content on Stack Overflow. In addition, we add to the academic and practitioner communities' understanding of regression approaches, in terms of the value they provide given the availability of particular data. The research questions that supported our portfolio of analyses is:

> *RQ1. What variables predict Stack Overflow code quality?*
>
> *RQ2. Which regression approach provides the best predictive power when modelling Stack Overflow code quality?*

The remaining sections of this study are organized as follows. In the next section we review the literature. We next provide our methods in Section III. Results are subsequently provided in Section IV. We next discuss these findings and outline implications in Section V, before considering threats to validity in Section VI. Finally, concluding remarks and future work are provided in Section VII.

## II. BACKGROUND AND RELATED WORK

### A. Exploring Public Software Code

While online platforms and other publicly available software code repositories like GitHub[1], GitLab[2] and Bitbucket[3] provide useful information for the practitioner community, the quality of code posted on these platforms has become an issue of contention. This is because most software developers depend on these platforms and code repositories for solutions [1, 2], and there has been failures and errors reported across the software development industry. For instance, the Ariane 5 Disaster failure [16] and Nissan ConnectEV error [3] are examples.

Thus, academic research has examined and reported on errors in code online. For instance, Meldrum et al. [1] have investigated the quality of Stack Overflow code in light of strong interest in this platform [18][14]. Source code quality was evaluated according to four aspects: reliability and conformance to programming rules, readability, performance, and security, where the authors found that Stack Overflow source code snippets contain on average 4.8 reliability and conformance to programming rules violations, 10.5 readability violations and 0.5 performance violations [1]. Fewer security violations were detected, notwithstanding that some serious security errors were seen [1]. Another study found that in Stack Overflow, there were no JavaScript code without code quality violations [4]. These authors checked 336,643 JavaScript code fragments from Stack Overflow using ESLint, where about 11.9 code violations for each code snippet on average were observed [4]. Further, Treude et al. [2] established that more than 50% of code snippets available on Stack Overflow are not self-explanatory. This is mainly because of incomplete code fragments, code quality issues, missing rationale, bad code organization, clutter, naming issues and missing domain information [2].

The use of these low-quality code snippets in software projects can result in inadequate software. For instance, Ahmad et al. [5] observed that code snippets copied from Stack Overflow into GitHub projects resulted in reduced code cohesion in over 70% of the cases. Others have attempted to predict Stack Overflow question quality. For instance, Baltadzhieva et al. [19] investigated mechanisms for predicting the quality of questions on Stack Overflow. They used question score and the number of answers received as the proxy measures for quality. The authors created four regression models using a mixture of variables where the model with the lowest mean absolute error was recorded. Variables that influenced the quality of questions were: the terms used to formulate the question, user reputation, the length of the question title and body, the presence of a code snippet and the question tags. The authors further explored the lexical terms that determine the quality of questions where they ranked the terms extracted from questions in Stack Overflow, finding several terms that improve or reduced quality. For instance, a range of markers negatively affected question quality, including terms which express excitement, negative experience or frustration, those expressing exceptions, if terms are expressed by a new user on Stack Overflow, if there is need for improvement of questions due to spelling mistakes or if there was off-topic questions and interjections [19]. Stack Overflow code was also found to be most change- and bug-prone when used in commercial projects [56]. On the backdrop that every day, new users, new questions and new answers are added to Stack Overflow, and the challenge created by the need to review and improve the quality of questions and answers on this platform, Tavakoli et al. [6] also proposed a novel method to improve the quality of the code snippets in answers on this platform. They created ExRec (Example Recommender), an eclipse plugin to extract code snippets into eclipse, refactor code to improve quality and upload the refactored code into Stack Overflow [6]. Datasets are also provided to encourage code quality experiments (e.g., [7]).

### B. Modelling Software Engineering Datasets

In supporting the data mining science, research has exploited several available software development datasets for studying the performance of various techniques and their ability to predict a range of issues. Malgonde et al. [8] proposed an ensemble-based model to predict the development effort of agile software development projects. They used support vector machine (SVM), k-nearest neighbors (KNN), artificial neural network (ANN), decision trees (DT), Ridge Regression, Linear Regression and Bayesian Network to create the ensemble-based model. Using the proposed model, they could predict software requirements effort and optimize sprint planning for two software projects [8]. To facilitate easy access to the most recent and best answers on Stack Overflow, Amancio et al. [9] proposed a ranking algorithm where recency and the quality of the answers are considered. They use textual and non-textual features extracted from users' answers where answers are grouped into high quality and low quality, using a threshold value. These groups are then sorted using the recency value, and joined to create a globally ranked list [9]. Using this ranking, developers can find the best answer for their issues. Shcherban et al. [10] used machine learning to identify posts related to code smells. They used Logistic Regression for their modelling where precision of 0.978, recall of 0.965 and F1-score of 0.971 were reported during evaluation [10]. With these results, it is indicated that machine learning can be used to identify questions with code smell discussions.

On the backdrop that 29% of the questions on Stack Overflow do not have an answer [11] and 47% of the questions are not resolved [20], Mondal et al. [11] tried to detect questions that were likely to go unanswered using four machine learning models, where 79% accuracy was achieved using the Random Forest approach. Yazdaninia et al. [20] also proposed to predict questions that are not likely to attract answers at the time of submission where several machine learning models were evaluated. Of the approaches used, XGBoost achieved 71% accuracy [20]. Tavakoli et al. [12] proposed to automatically identify deficient Stack Overflow posts where the J48 decision tree algorithm was used to achieve 94.5% precision and 90.3% recall. Toth et al. [13] used deep learning and natural language processing (NLP) techniques to train a model with linguistic and semantic features to detect Stack Overflow posts for closure, where 74% accuracy was reported. Singh et al. [21] tried to predict Stack Overflow post tags where a precision of 0.75, recall of 0.391 and F1 score of 0.523 were returned by TagAssc. Related work was done by Ahasanuzzaman et al. [22] to identify posts related to a specific category. A supervised learning approach called CAPS was developed using Conditional Random Fields to identify posts related to API issues. These authors achieved values of 0.95 and 0.71 for precision and recall for the issue class and 0.76 and 0.95 precision and recall for the non-issue class [22]. More recent efforts also aim to detect and repair faults in Stack Overflow code, with promising results [57].

---

[1] https://github.com/

[2] https://about.gitlab.com/

[3] https://bitbucket.org/

While various studies have explored the quality of code online and investigated various mechanisms for predicting content on these platforms, less interest is placed in evaluating techniques for the conditions in which they were designed for use. We fill this gap in this study by first exploring the variables that predict Stack Overflow code quality before evaluating the performance of various regression-based approaches in predicting code quality.

### III. METHODS

#### A. Data Acquisition and Processing

A similar methodology to Meldrum et al. [1] was adopted to acquire the data and to prepare it for input to the various regression approaches. Data was sourced from Stack Overflow for 2014, 2015, and 2016 using Stack Overflow's data explorer[4] via standard SQL statements, not only as these years had the highest number of questions and answers on Stack Overflow but also a prior study had confirmed a good level of code reuse evidenced by the software engineering community for these snippets [23].

Java code were sampled due to Java's popularity with the expectation that the sample would provide transferable insights to Java code on Stack Overflow for the software development community given Stack Overflow data are generalizable across languages and time [23][24]. Answer posts which contained at least one "<code>" tag and were from a question tagged as Java were then sampled. As Stack Overflow's answers are structured in HTML style with code snippets included between code tags (i.e., <code> </code>), this structure allowed for appropriate identification of relevant code snippets as these "<code>" tags would indicate the presence of a code snippet. The resulting data set comprised of 117,523 answers (46,103 accepted answers and 71,423 unaccepted answers). Other attributes relating to the questions that generated these answers were also extracted and the entire data set imported into a Microsoft SQL server database.

Using a small Java program, the Java code was then extracted from the *answerBody*, with each code snippet between the tags "<code>. . .</code>" being saved separately. Analysis of this Stack Overflow code revealed two types of code snippets: 1) in-line code snippets (surrounded by <code> tags) which were typically added as part of text answers when contributors needed to provide explanation, and 2) code blocks (surrounded by <pre><code> tags) which provided more substantial implementation of specific solutions. Lotter et al. [23] argued for the exclusion of these in-line code snippets to be analyzed for quality as they felt it to be unrealistic and unfair, and thus these were removed from our sample. This left 404,779 code snippets, from the 117,526 answer posts (191,556 snippets from accepted answers and 213,223 snippets from unaccepted answers).

Next, only code snippets with one or more lines of code (i.e., contained a new line character) were analyzed for quality in accordance with previous recommendations ([23][25][26]). For example, code snippets that did not contain a new line character such as those single word Java code snippets were excluded. This was achieved by extracting snippets from the database using a Java application which checked if the code snippets contained 'import', 'package', or 'class' and then saving the file unmodified. Code snippets that did not contain these words were encased in a public class structure as described in Yang et al. [26]. Each file was saved as a .java file with a unique identifier (e.g., C1234.java represented code snippet number 1234) so that it was possible to trace each code snippet. There were minimal repairs on code snippets to reduce the risk of confounding the results and analysis. The individual .java files are available as part of our replication package here: https://tinyurl.com/uv8jtfq.

In total, 151,954 of the code snippets from 94,279 answers were used for analyses (i.e., 37.5% of the code snippets). This sample comprises 66,389 snippets from accepted answers and 85,565 snippets from unaccepted answers, which allowed the quality of snippets to be compared across these two groups. Attributes included in this dataset are answer identification number (*answerId*), question identification number (*questionId*), answer score (*answerScore*), answer creation date (*answerCreationDate*), answer body (*answerBody*), question date (*questionDate*), question score (*questionScore*), view count (*ViewCount*), answer count (*AnswerCount*), and comment count (*CommentCount*). LOC[5], Code Length[6], Code Space[7], and SPA[8] were also computed to aid the analyses. Fields unrelated to code that were largely 'null', for example history logs, were omitted. The majority of code snippets were between 4 and 60 LOC.

Tool selection to analyze the code snippets for these quality attributes was based on the work of Meldrum et al. [1] who argued for using a combination of PMD[9], CheckStyle[10], and FindBugs[11] as these tools have been effective in analyzing code snippets in the past. Moreover, each tool has complementary categories of checks which allow for greater quality coverage of the code snippets than using one tool alone [1]. FindBugs was able to check for security and malicious code vulnerabilities. Both PMD and CheckStyle were applied to check for performance. Readability and compliance with Google coding standards were checked using CheckStyle. The full list of checks configured are described by Meldrum et al. [1]. We adopted their approach given the comprehensive review of various code quality techniques and tools and reliability checks that were done before specific metrics were proposed.

---

[4] https://data.stackexchange.com/stackoverflow/query/new

[5] Refers to the number of lines of code.

[6] Refers to the length of the code snippet in terms of number of characters.

[7] Refers to the number of space characters (i.e., ' ') in the code snippet provided as a part of answers.

[8] Refers to the number of code snippets per answer associated with each code snippet.

[9] https://pmd.github.io.

[10] http://checkstyle.sourceforge.net.

[11] http://findbugs.sourceforge.net/index.html.

For instance, Fig. 1 provides a sample reliability and conformance to programming rules error as detected by PMD, where code at line 14 should be replaced with a subclass exception to fix the error (e.g., NoSuchField Exception). In Fig. 2 a sample readability error as returned by CheckStyle is provided where there is no code included in the catch block at line 27, violating the Empty Catch Block rule. Fig. 3 provide a sample performance error as returned by FindBugs where an Inefficient Number Constructor is used at line 7. Here the code performance would improve if the code at line 7 was replaced with 'Double.valueOf(amount + 10)'. In Fig. 4 several 'public static' variables are used inappropriately, as detected by FindBugs security checks. These fields should be 'final' (code lines 2-4 and 6). It was common for snippets to return multiple violations, at times even for a single code quality dimension (e.g., for Performance). Our raw snippet dataset and processed code quality outcomes are available as part of the replication package here: https://tinyurl.com/uv8jtfq.

Fig. 1. Sample PMD reliability and conformance to programming rules error from Stack Overflow code

Fig. 2. Sample CheckStyle readability error from Stack Overflow code

### B. Competing Regression Algorithms and Measures

A common problem is to adequately approximate a function of several variables requiring an approach to model a response variable, $y$, on one or more predictor variables $x_1, \ldots, x_n$ given a set of data. The underlying assumption is that the system that generated the data is of the form $y = f(x_1, \ldots, x_n) + \varepsilon$ [27], where $\varepsilon$ is an additive stochastic component which is neither controlled for nor observed [28]. Approaches to construct this function fall into the domain of regression analysis where the data is used to construct a function (model) $\hat{f}(x_1, \ldots, x_n)$ which would act as a reasonable approximation of $f(x_1, \ldots, x_n)$ [28]. Instead of adopting a single regression approach to develop a model, six different regression approaches were used to generate models based on the output of the tools' that were used to analyze the code snippets in following recommended guidelines [29]. It is important to study these approaches as some of these have a long history of adoption and others are based on more recent approaches for generating better approximations for the models. Furthermore, some regression approaches deal with non-normal distributions of the response variable better than other methods.

Fig. 3. Sample FindBugs performance error from Stack Overflow code

Fig. 4. Sample FindBugs security error from Stack Overflow code

The simplest regression approach is that of **Linear Regression** which uses ordinary least squares to estimate the parameters of the model. As it is a classical approach, it is considered to be a baseline model that other sometimes more complex approaches are compared with [15]. However, adopting a linear model is contingent on the following assumptions. That for each value of the predictor variable, the distribution of the response variable must be normal. All observations should be independent and there is a linear relationship between the response variable and the predictor variables [15]. Finally, all values of the predictor variable should be constant with respect to the variance of the distribution of the response variable [15].

When the distribution of the response variable is not normal, a different approach, **Generalized Linear Models (GLM)**, can be applied. This linear model is augmented with a link function that generalizes it to allow the magnitude of the variance of each measurement to be a function of its predicted value [30]. The default method proposed in Nelder and Wedderburn [30] instead employs an iteratively reweighted least squares method to estimate the model's parameters for maximum likelihood. This results in generating a model with a high probability of it fitting the data [30].

An alternative approach is to use **Classification and Regression Trees (CART)** [31] as opposed to a least squares method to estimate the parameters of the model for the entire data space. It is obtained by recursively partitioning the data space into subregions and fitting a constant estimate of $y$ within each subregion [31]. Once the tree has been generated, it can be depicted as a decision tree. Unlike Classification Trees where the response variable can take a finite set of unordered labels, Regression Trees specifically are for response variables that take continuous values [31]. However, the choice around the resulting depth or size of the tree impacts on the performance of it. Too shallow and the tree underfits the training data, and so does a poor job at generalizing to new data, whereas large trees tend to overfit the data [32]. Over the years different tree pruning strategies have been proposed to overcome these weaknesses [32].

Another main criticism of Classification and Regression Trees is that the product of recursively partitioning the data space results in the approximating function being discontinuous at the boundaries of the subregions that limits its effectiveness [28]. To overcome this limitation, Friedman [28] proposed a combination of Generalized Additive Models where another function is used to capture (smooth) the important parts of the data and Regression Trees. Friedman [28] refers to this as the **Multivariate Adaptive Regression Spline (MARS)** which adopts an expansion of product spline functions that models non-linear data and interactions among inputs. MARS (implemented as **Earth** due to trademark considerations) automatically determines the spline number and parameters from the data using recursive partitioning but distinguishes between the additive contributions of each input and interactions among them [33]. Iteratively adding the functions reduces the residual until a stopping criterion is met [33].

However, a key assumption underlying the approaches described above is that one model is generated but there are advantages to generating an ensemble of weak models which when combined forms the prediction model. This has the benefit of increasing predictive performance through diversity in the models [34] which are typically decision trees [35]. The **Gradient Boosting Machine (GBM)** uses ensemble learning with each successive tree, learning and improving on the previous tree and generating gradient boosted trees which finds the optimal model to make predictions with the given data by minimizing a loss function [36]. More recently, **XGBoost**, a popular GBM-based algorithm offers extensions to GBM to reduce the model overfitting to the data and lowering the complexity of the functions generated by each tree whilst still preserving the predictive accuracy of the model [37].

There is also the need to evaluate the regression approaches applied in our work for predictive power. That is, how well is the discrepancy between the observed values and those produced by the model which is also known as the goodness-of-fit [38]. Widely used metrics are the coefficient of determination ($R^2$) or the residual variance (error mean square $\sigma_\varepsilon^2$) [27]. Not doing this means it is difficult to make any claim as to how confident we are as to whether a model can generate reasonable predictions or not. More importantly, such performance metrics allow for more measurable and objective means for assessing the predictive power of a regression approach [38]. Finally, appropriate performance metrics should be selected to assess an individual model's performance and to compare different models to establish the best performing model. The Root Mean Square Error (RMSE) [39], for example, is a frequently used metric to compare the predictive performance across models and we use such a metric in the work reported in this study.

*C. Reproducibility Checklist and Parameters*

For implementations of the regression approaches, *xgbTree* and *cart* functions were used from the R caret package, the *earth* function was used from the R earth package which is an implementation of the MARS method, and the R *gbm* library provisioned the gbm function. The other two approaches, *linear* and *glm* functions, were part of the R standard library. In facilitating easy reproducibility of our outcomes, we note that hyper-parameter optimization were used to establish the best learning parameters of the R implementations for the regression algorithms. More specifically, the same random seed was used for all folds for cross validation to ensure consistency across the models generated. For each regression approach we generated 20 models and selected the model that obtained the lowest RMSE. We then repeated 10-fold cross validation on each of these optimal models twice, producing 20 results per model. Different random seeds were used to generate the folds for cross validation. For example, the optimal learning parameters for gbm were num_trees = 1843, interaction_depth = 6, shrinkage = 0.1223 and min_observations_in_node = 5. All data sets, generated data and scripts are provided in the replication package available here: https://tinyurl.com/uv8jtfq. The computer used for our experiments had an Intel Core i9 processor, 32GB of RAM with 1TB storage capacity. We will provide a permanent link to our replication package on Zenodo after the manuscript is reviewed, in compliance with the blind review process.

IV. RESULTS

We recorded 244,266 reliability and conformance to programming rules errors (mean=4.82), 530,521 readability errors (mean=10.51), 3,949 performance errors (mean=0.49) and 75 security errors (mean=0.01). These errors were found in 50,717 Java code snippets. Here, like others (e.g., [4]), we see that Stack Overflow code contained violations, and particularly those related to reliability and conformance to programming rules, readability and performance. Given our goal to examine the variables that predict Stack Overflow code quality, and that some of our data violated normality, we performed Kendall Rank correlation analysis to understand how the variables interacted and informed our modelling. In terms of high effect size, Table I shows that code length correlated with code spaces and lines of code, and this variable also correlated with Reliability and Conformance to Programming Rules and Readability violations. A similar pattern of outcome was observed for code spaces and lines of code. In addition, code snippets that suffered Reliability and Conformance to Programming Rules violations also returned many Readability violations. Other notable correlations (medium effect size) include question score and view count, and answer score and accepted answer. These were all statistically significant results ($p < .05$).

We next examine the specific variables that predict Stack Overflow code quality. Given that the outcome variable (Number of Violations) was numeric, as noted above, we used regression modelling. We first explored the literature for recommendations on baseline models. Here we observed that several approaches are used across domains, with Luz et al. [40] recommending numerous approaches, including linear regression as a baseline model to provide a solution for the

ADReSS Challenge at Interspeech 2020. Kitchenham et al. [41] have done a literature review on different approaches used for cost estimation. In that paper, both the cross-company cost estimation models and within company cost estimation approaches were studied. Outcomes reported by the authors established that linear regression is the most suitable baseline model [41]. Lenberg et al. [42] used linear regression to analyze software engineers' attitudes towards organizational change. Notwithstanding the repeated use of linear regression for baseline modelling, Whigham et al. [43] noted that skewness in data can affect linear models' performance, and thus, they recommend automatically transformed linear model (ATLM) as the best baseline model. We thus used Whigham et al.'s [43] approach to generate our baseline model. We created a 15[th] column in our dataset named 'Total Violations', which provided the sum of all violations, before we transformed our dataset. Following Whigham et al. [43], we checked each column for the violation of normality assumption (via skewness), and used a log transformation if the data was not normal. Whenever this procedure did not result in normality, we instead used a square root transformation. Columns that demonstrated a normal distribution already were left unchanged (see 'Measure' column in Table I with indexes included for the different transformations). We then performed a multiple regression, where the code attributes were regressed against the total violations, with outcomes showing that the model explained 81% of the variance in code snippet quality and was significant $F(10,50706) = 22042.17$, $p < 0.01$.

In terms of the weight of the code attributes, Table II shows that answer count (Answer Count), code length (Code Length), code spaces (Code Spaces) and lines of code (LOC) were most influential (with coefficients 0.138480, 0.382147, 0.099801 and 1.958790 respectively). We observed that code length, code spaces and lines of code significantly correlated with the presences of violations above, however, answer count did not have such a presence in the correlation analysis.

We next investigate xgbTree, earth, gbm, linear (lm), glm and cart for their predictive power. This analysis is done to explore the regression approach that provides the best predictive power. To establish optimal learning parameters for the models we adopted 10-fold cross validation using the appropriate R package (refer to Section III.C for details) [45]. We plot the RMSE, the standard deviation of the residuals, which reveals how data points are clustered around the line of best fit. A lower RMSE is favored, and signals better performing models [46, 47]. Fig. 5 shows that gbm recorded the best mean performance (RMSE=2.77), while lm recorded the worst (RMSE=40.55).

TABLE I. KENDALL RANK CORRELATION (T) ANALYSIS FOR CODE ATTRIBUTES AND VIOLATIONS

| | Measure | 1 | 2 | 3 | 4 | 5 | 6 | 7 | 8 | 9 | 10 | 11 | 12 | 13 | 14 |
|---|---|---|---|---|---|---|---|---|---|---|---|---|---|---|---|
| Code Attributes | 1. Question Score[n] | 1 | *0.33* | *0.14 | *-0.03 | *0.28 | *0.03 | *0.03 | *0.03 | *0.08 | *-0.03 | 0.00 | 0.00 | 0.01 | 0.00 |
| | 2. View Count[l] | | 1 | *0.20 | *0.10 | *0.12 | *0.15 | *0.15 | *0.14 | 0.00 | *-0.06 | *0.12 | *0.12 | *0.03 | -0.02 |
| | 3. Answer Count[l] | | | 1 | *0.11 | *-0.05 | 0.01 | *0.03 | *0.02 | *-0.09 | *-0.27 | *0.04 | *0.03 | 0.00 | -0.02 |
| | 4. Comment Count[s] | | | | 1 | 0.00 | *0.07 | *0.09 | *0.09 | 0.00 | -0.02 | *0.08 | *0.08 | 0.01 | 0.00 |
| | 5. Answer Score[n] | | | | | 1 | *-0.10 | *-0.09 | *-0.09 | *0.28 | *0.34* | *-0.09 | *-0.13 | *0.03 | -0.01 |
| | 6. Code Length[l] | | | | | | 1 | **\*0.96** | **\*0.94** | *-0.27 | 0.00 | **\*0.83** | **\*0.79** | *0.10 | *-0.03 |
| | 7. Code Spaces[l] | | | | | | | 1 | **\*0.95** | *-0.26 | 0.00 | **\*0.82** | **\*0.82** | *0.10 | *-0.04 |
| | 8. LOC[l] | | | | | | | | 1 | *-0.25 | 0.01 | **\*0.83** | **\*0.82** | *0.09 | *-0.04 |
| | 9. SPA[l] | | | | | | | | | 1 | *0.24 | *-0.25 | *-0.27 | 0.00 | 0.02 |
| | 10. Accepted (1) /Unaccepted (0)[n] | | | | | | | | | | 1 | -0.01 | -0.01 | 0.02 | -0.01 |
| Number of Violations[s] | 11. Reliability and Conformance to Programming Rules[s] | | | | | | | | | | | 1 | **\*0.73** | *0.16 | *-0.04 |
| | 12. Readability[s] | | | | | | | | | | | | 1 | *0.11 | *-0.04 |
| | 13. Performance[s] | | | | | | | | | | | | | 1 | -0.02 |
| | 14. Security[s] | | | | | | | | | | | | | | 1 |

Keys: * indicates significance (p < 0.05), *italics* indicates medium effect size using Cohen's [44] classification (0.3 ≤ rho ≤ 0.49), **bold** indicates high effect size using Cohen's classification (rho ≥ 0.50), n=normal distribution, l=log transformation, s=square root transformation

TABLE II. REGRESSION COEFFICIENTS OF SIGNIFICANT CODE ATTRIBUTES

| Code Attributes | Coefficient |
|---|---|
| Question Score | -0.001257 |
| View Count | -0.037481 |
| Answer Count | 0.138480 |
| Comment Count | 0.041386 |
| Answer Score | -0.001242 |
| Code Length | 0.382147 |
| Code Spaces | 0.099801 |
| LOC | 1.958790 |
| SPA | -0.063379 |
| Accepted | -0.032957 |

We further examine these outcomes for statistical differences using a Kruskal-Wallis H test with Bonferroni adjustments for post hoc pairwise comparisons. Findings confirmed statistical differences across the six modelling approaches, $(X^2(5) = 104.70, p < 0.01)$, with pairwise comparisons results in Table III also confirming the patterns of outcomes in Fig. 5. Table III shows that gbm, xgbTree and cart performed significantly better than the three other approaches, with gbm returning the best performance overall. The linear model performed similar to glm. As can be seen in Table III not all patterns of outcomes are statistically significant (p<0.01), but it is clear that glm and linear are the least performing models compared with gbm, xgbTree, and cart, with earth exhibiting performance between that of all the other models. We validate these outcomes in Section V.C.

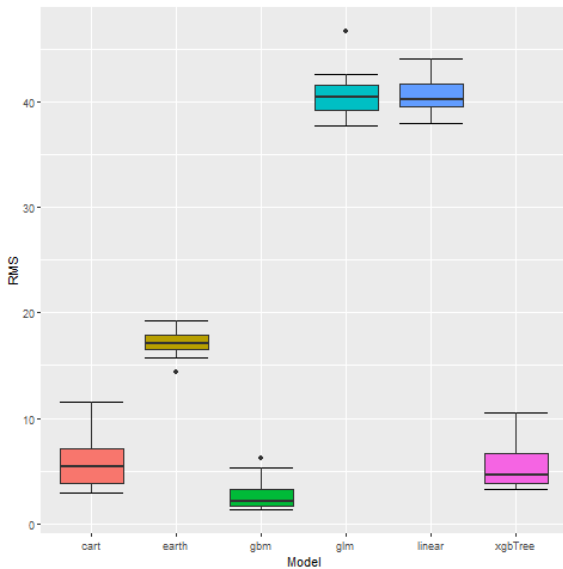

Fig. 5. RMSE outcomes for Code Quality regression models

TABLE III. KRUSKAL-WALLIS RESULTS FOR PAIRWISE COMPARISONS

| Sample 1-Sample 2 | Test Statistic | Bonferroni Adjusted Significance |
|---|---|---|
| gbm-xgbTree | 1.9092 | 0.8436 |
| gbm-earth | -4.9865 | 0.0000 |
| gbm-linear | 7.7139 | 0.0000 |
| gbm-glm | 7.7139 | 0.0000 |
| gbm-cart | -2.1410 | 0.4841 |
| xgbTree-earth | -3.0774 | 0.0313 |
| xgbTree-linear | -5.8047 | 0.0000 |
| xgbTree-glm | -5.8047 | 0.0000 |
| xgbTree-cart | -0.2318 | 1.0000 |
| earth-linear | 2.7274 | 0.0958 |
| earth-glm | 2.7274 | 0.0958 |
| earth-cart | 2.8456 | 0.0665 |
| linear-glm | 0.0000 | 1.0000 |
| linear-cart | 5.5729 | 0.0000 |
| glm-cart | 5.5729 | 0.0000 |

## V. DISCUSSION AND IMPLICATIONS

We answer our research questions *RQ1. What variables predict Stack Overflow code quality?* and *RQ2. Which regression approach provides the best predictive power when modelling Stack Overflow code quality?* in this section. We first examine the variables that predict Stack Overflow code quality, before then teasing out the regression approach with the best predictive power.

### A. Variables that Predict Stack Overflow Code Quality

Our model only explained 81% of the variance in code snippet quality, so we concede that there are other important features missing from our data set. However, it is important to note that longer Stack Overflow code tended to have more violations. This finding is consistent with those reported by others (e.g., [1]). It seems that with the opportunity to provide more insights for the community, contributors on Stack Overflow seem to attempt to do so without quality considerations at times. For instance, it was common to see a range of reliability and conformance to programming rules errors that were both serious and trivial. A specific serious error evident in longer code was the throwing of 'Raw Exception Types'. This error causes confusion for programmers, where it is unclear to understand the reason for faults, making it hard to debug the code.

We observed that questions that were scored higher also attracted more views. This is fitting, and perhaps a better assessment of potential quality than the total number of answers that are recorded [19]. The community here seems to pay more attention to higher quality solutions on Stack Overflow. In fact, the community also tended to use accepted answers as a basis to judge quality, as our outcomes show that accepted Stack Overflow answers tended to score higher than those that were not accepted. From a question perspective, it was previously observed that the quality of questions were influenced by the terms used to formulate the question, user reputation, the length of the question title and body, the presence of a code snippet and the question tags [19]. That said, these authors used question score and the number of answers received as the proxy measures for quality, which are not as objective as the measures we have used, where actual errors present in code snippets are evaluated.

In fact, the more answers that are added to questions the more errors were typically observed in the code that was provided in these answers. This outcome goes against the value of online portals and open communities more generally, where the power of the crowd is said to act to enforce quality. Here we see that quality reduced as answers are added, which may not be ideal for the increasing reuse of content on Stack Overflow platform [14], where errors can propagate and be lasting [5]. Could it be that those providing new solutions are less prudent knowing that others have already tackled the issue identified/requested in the question? If this is the case, the motivation to add additional solutions/answers with poor quality seems unusual. Contributors' behavior seems to be at odds here, and points to code quality issues potentially being influenced by more than the lack of explanation [2], tending towards bad habits being disseminated in the code contributed itself. While mechanisms that allow others to provide corrections could be helpful [6] (including in future comments provided for posts), at the core of quality improvement would need to be a cultural shift in behavior.

### B. Regression Approach with Best Predictive Power

In terms of the regression models, we observed that Gradient Boosting Machine (gbm) performs the best of all regression approaches that were investigated in this study (including linear regression, xgbTree, glm, earth and cart) for their predictive power. Gradient Boosting Machine uses ensemble learning where each successive tree learns and improves on the previous tree generating gradient boosted trees which finds the optimal model to make predictions with the given data by minimizing a loss function [36]. The low RMSE of gbm could be explained by the fact that the advantages of gradient boosting is to improve model accuracy whilst at the same time performing variable selection and model choice [48]. Likewise, as gbm is based on a greedy additive algorithm it will not select at any step any variable that is redundant, thereby reducing the effect of collinearity [36]. Our outcomes here show this learning process enhanced the prediction accuracy of the model. In fact, XGBoost previously achieved 71% accuracy in predicting questions that are not likely to attract answers at the time of submission, which was superior to other approaches used [20]. This is a popular GBM-based algorithm which offers extensions to GBM to reduce the model overfitting of the data and lowering the complexity of the functions generated by each tree whilst still preserving the predictive accuracy of the model [37]. The lower performance of XGBoost compared with gbm could have been attributed to the level of regularization specified to prevent overfitting of the model [37]. We observed that cart had the next best performance compared to gbm and XGBoost. It is possible that the performance of cart was affected by the choice around the resulting depth or size of the tree potentially impacting on model performance [32]. The performance of earth may be explained by the procedure used for stepwise selection of the variables as pointed out by others [49], where it was noted that these approaches would not guarantee the best set of variables for a given size. Furthermore, earth is likely to overfit the data when there are a large number of predictor variables and there is no precise information about how these variables are related exactly [50]. The outcomes from the other approaches (linear and glm) were unremarkable.

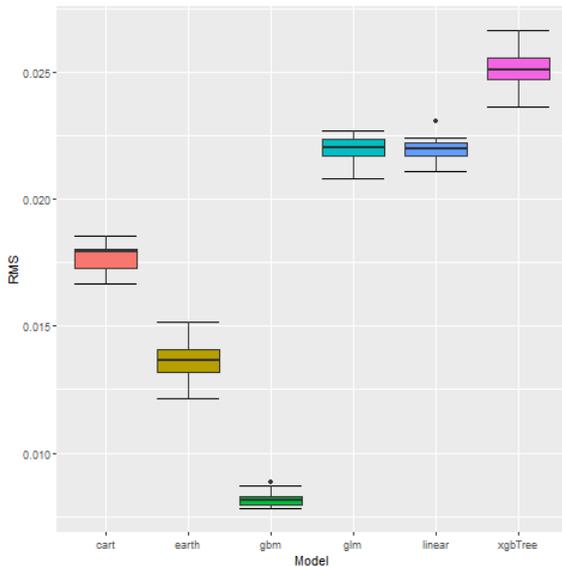

Fig. 6. RMSE outcomes for the Electrical Grid Stability Simulated dataset

### C. Regression Approaches Validation

We validate the pattern of outcomes by applying our methodology to two different independent regression data sets to assess the aforementioned six models' performance. The identical process was followed, with all data sets, generated data and scripts provided in the replication package available here: https://tinyurl.com/uv8jtfq. The Electrical Grid Stability Simulated dataset [12] contained 14 attributes with 10,000 instances and the Cancer Mortality Dataset for USA Counties[13] contained 34 attributes with 3,048 instances. With the latter dataset, two columns were removed as these contained categorical values. As can be seen in Fig. 6 and Fig. 7, gbm was the best performing model in both instances. Outcomes here supports the position that our methodology is generalizable to different regression data sets and gbm continues to maintain superior performance over the other five approaches. The pattern of outcomes in Fig. 6 and Fig. 7 are not consistent for the other regression approaches however. Although the RMSE for cart and xgbTree recorded among the top three outcomes in Fig. 6 and Fig. 7 respectively, earth also saw slight improvement (2nd and 3rd best performance) when compared to its 4th place performance in Fig. 5.

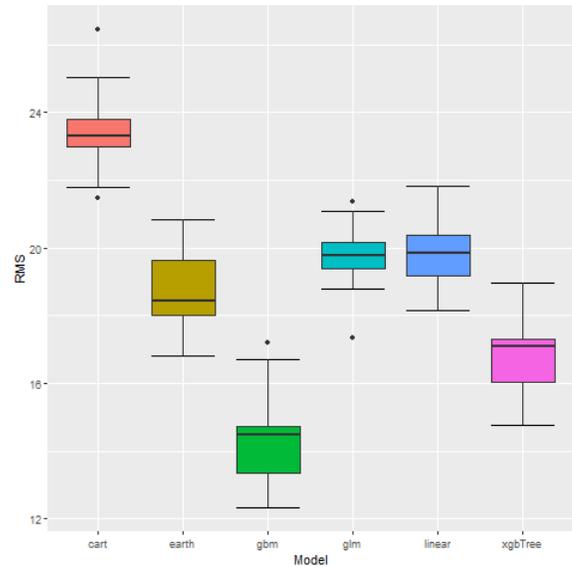

Fig. 7. RMSE outcomes for the Cancer Mortality for USA Counties dataset

## VI. THREATS TO VALIDITY

Our code data set only comprised Java-based artifacts from Stack Overflow, which affects the generalizability of our outcomes. While we concede this limitation, some of the outcomes in our work are similar to those that previously analyze other code [23, 51, 52], suggesting that some of the errors observed may be relevant in other contexts. The tools that are used in this work may possess limitations, albeit these tools have been accepted by the academic and practitioner communities [1, 53]. For instance, Ayewah and Pugh [54] evaluated FindBugs using a dataset from Google, finding that this tool was very effective at detecting bugs, and its use was considered to be beneficial for saving time and money for developers. Rigorous evaluations previously done on the outcomes of the tools by others have also reported accurate outcomes [1]. We concede that we have examined a limited set of features, and the textual features in the post itself may be noteworthy in helping to predict code quality. Finally, we have investigated the performance of regression approaches in this study, and thus, we cannot definitively say that our outcomes will hold for other modelling methods (e.g., Random Forest or SVM). While this may be considered a limitation, our work adds to the body of knowledge where there is scarcity of details around the performance of various modelling approaches. We have also independently validated Gradient Boosting Machine (gbm) to be the best regression approach. We thus believe that notwithstanding the threats to validity of this study, our outcomes would be useful for the practitioner and academic communities.

---
[12]https://archive.ics.uci.edu/dataset/471/electrical+grid+stability+simulated+data

[13]https://github.com/DeepanshuManchanda/Predict-cancer-mortality-rates-for-US-counties.-Exp-1-2

VII. CONCLUSION AND FUTURE WORK

In adding to the body of evidence around software quality on online portals and other publicly available software code repositories and the evaluation of approaches that facilitate modelling and prediction of various software engineering issues, we examined the variables that predict Stack Overflow code quality, and the regression approach that provides the best predictive power. We observed that longer Stack Overflow code tended to have more violations, questions that were scored higher also attracted more views and the more answers that are added to questions the more errors were typically observed in the code that was provided in these answers. In addition, Gradient Boosting Machine (gbm) performed the best of all regression approaches that were investigated in this study (including linear regression, xgbTree, glm, earth and cart). These outcomes show that there is scope for the Stack Overflow community to improve and adopt a cultural shift in their behavior towards sharing high quality artifacts. Further, the mechanisms used by gbm (e.g., gradient boosting) are noteworthy in the way they improve prediction outcomes when compared to other regression approaches.

Our future work will aim to understand the specific reasons for the way code quality on Stack Overflow degrades when code was longer or more snippets were added by the community. In addition, we hope to investigate mechanisms for further optimizing the regression models to achieve better predictive performance. For example, with cart there are a number of existing pruning strategies that can be applied [31]. In addition, examining the regression models to establish variable importance through the likes of Shapley Values [55] will provide other insight into specific variables that significantly influence Stack Overflow Code quality. Finally, we seek to understand if and how variable importance changes over time and what effect it has on the regression models' performance.

REFERENCES

[1] . Meldrum, S. A. Licorish, C. A. Owen, and B. T. R. Savarimuthu, "Understanding stack overflow code quality: A recommendation of caution," *Science of Computer Programming,* vol. 199, pp. 102516-102516, 2020, doi: 10.1016/j.scico.2020.102516.

[2] C. Treude and M. P. Robillard, "Understanding stack overflow code fragments," *Proceedings of IEEE Int. Conf. on Software Maintenance and Evolution,* pp. 509-513, 2017, doi: 10.1109/ICSME.2017.24.

[3] "Nissan app developer busted for copying code from Stack Overflow - The Verge." https://www.theverge.com/tldr/2016/5/4/11593084/dont-get-busted-copying-code-from-stack-overflow (accessed 15th May, 2021).

[4] U. Ferreira Campos, G. Smethurst, J. P. Moraes, R. Bonifacio, and G. Pinto, "Mining rule violations in javascript code snippets," *IEEE International Working Conference on Mining Software Repositories,* vol. 2019-May, pp. 195-199, 2019, doi: 10.1109/MSR.2019.00039.

[5] M. Ahmad and M. O. Cinneide, "Impact of stack overflow code snippets on software cohesion: A preliminary study," *IEEE International Working Conference on Mining Software Repositories,* vol. 2019-May, pp. 250-254, 2019, doi: 10.1109/MSR.2019.00050.

[6] M. Tavakoli, A. Heydarnoori, and M. Ghafari, "Improving the quality of code snippets in stack overflow," presented at the Proceedings of the 31st Annual ACM Symposium on Applied Computing, 2016.

[7] V. Lenarduzzi, N. Saarimäki, and D. Taibi, "The technical debt dataset," in *Proceedings of the Fifteenth International Conference on Predictive Models and Data Analytics in Software Engineering*, 2019, pp. 2-11.

[8] O. Malgonde and K. Chari, "An ensemble-based model for predicting agile software development effort," *Empirical Software Engineering,* vol. 24, no. 2, pp. 1017-1055, 2019.

[9] L. Amancio, C. F. Dorneles, and D. H. Dalip, "Recency and quality-based ranking question in CQAs: A Stack Overflow case study," *Information Processing & Management,* vol. 58, no. 4, p. 102552, 2021.

[10] S. Shcherban, P. Liang, A. Tahir, and X. Li, "Automatic Identification of Code Smell Discussions on Stack Overflow," presented at the Proceedings of the 14th ACM / IEEE International Symposium on Empirical Software Engineering and Measurement (ESEM), 2020.

[11] S. Mondal, C. M. K. Saifullah, A. Bhattacharjee, M. M. Rahman, and C. K. Roy, "Early Detection and Guidelines to Improve Unanswered Questions on Stack Overflow," presented at the 14th Innovations in Software Engineering Conference (formerly known as India Software Engineering Conference), 2021.

[12] M. Tavakoli, M. Izadi, and A. Heydarnoori, "Improving Quality of a Post's Set of Answers in Stack Overflow," presented at the 2020 46th Euromicro Conference on Software Engineering and Advanced Applications (SEAA), 2020.

[13] T. Gyimóthy, L. Vidács, D. Janthó, B. Nagy, and L. Tóth, "Towards an Accurate Prediction of the Question Quality on Stack Overflow using a Deep-Learning-Based NLP Approach," presented at the Proceedings of the 14th International Conference on Software Technologies, 2019.

[14] S. Meldrum, S. A. Licorish, B. Tony, and R. Savarimuthu, "Exploring Researchers ' Interest in Stack Overflow : A Systematic Mapping Study and Quality Evaluation," pp. 1-38, 2020. [Online]. Available: https://arxiv.org/abs/2010.12282.

[15] E. C. Alexopoulos, "Introduction to multivariate regression analysis," (in eng), *Hippokratia,* vol. 14, no. Suppl 1, pp. 23-28, 2010. [Online]. Available: https://pubmed.ncbi.nlm.nih.gov/21487487/.

[16] "The Worst Computer Bugs in History: The Ariane 5 Disaster | Bugsnag Blog." https://www.bugsnag.com/blog/bug-day-ariane-5-disaster (accessed 5th June 2021).

[17] C. Mayhew and M. D. Ahmed, "Software Implementation - Lessons to be learnt from the Novopay project," *26th Annual Conference of Computing & Information Technology Research & Education New Zealand,* no. October, pp. 1-7, 2013.

[18] A. Tahir, J. Dietrich, S. Counsell, S. Licorish, and A. Yamashita, "A large scale study on how developers discuss code smells and anti-pattern in Stack Exchange sites," *Information and Software Technology,* vol. 125. March, pp. 106333-106333, 2020, doi: 10.1016/j.infsof.2020.106333.

[19] A. Baltadzhieva and G. Chrupała, "Predicting the quality of questions on stackoverflow," in *Proceedings of the international conference recent advances in natural language processing*, 2015, pp. 32-40.

[20] M. Yazdaninia, D. Lo, and A. Sami, "Characterization and Prediction of Questions without Accepted Answers on Stack Overflow," *arXiv preprint arXiv:2103.11386,* 2021.

[21] P. Singh, R. Chopra, O. Sharma, and R. Singla, "Stackoverflow tag prediction using tag associations and code analysis," *Journal of Discrete Mathematical Sciences and Cryptography,* vol. 23, no. 1, pp. 35-43, 2020, doi: 10.1080/09720529.2020.1721857.

[22] M. Ahasanuzzaman, M. Asaduzzaman, C. K. Roy, and K. A. Schneider, "CAPS: a supervised technique for classifying Stack Overflow posts concerning API issues," *Empirical Software Engineering,* vol. 25, no. 2, pp. 1493-1532, 2019, doi: 10.1007/s10664-019-09743-4.

[23] A. Lotter, S. A. Licorish, B. T. R. Savarimuthu, and S. Meldrum, "Code Reuse in Stack Overflow and Popular Open Source Java Projects," in *2018 25th Australasian Software Engineering Conference (ASWEC)*, 26-30 Nov. 2018 2018, pp. 141-150, doi: 10.1109/ASWEC.2018.00027.

[24] D. Yang, P. Martins, V. Saini, and C. Lopes, "Stack Overflow in Github: Any Snippets There?," in *2017 IEEE/ACM 14th International Conference on Mining Software Repositories (MSR)*, 20-21 May 2017 2017, pp. 280-290, doi: 10.1109/MSR.2017.13.

[25] M. Duijn, A. Kucera, and A. Bacchelli, "Quality Questions Need Quality Code: Classifying Code Fragments on Stack Overflow," in *2015 IEEE/ACM 12th Working Conference on Mining Software Repositories*, 16-17 May 2015 2015, pp. 410-413, doi: 10.1109/MSR.2015.51.

[26] D. Yang, A. Hussain, and C. V. Lopes, "From query to usable code: an analysis of stack overflow code snippets," presented at the Proceedings of the 13th International Conference on Mining Software Repositories, Austin, Texas, 2016. doi: 10.1145/2901739.2901767.

[27] Y. T. Prairie, "Evaluating the predictive power of regression models," *Canadian Journal of Fisheries,* vol. 53, no. 3, pp. 490-492, 1996.

[28] J. H. Friedman, "Multivariate Adaptive Regression Splines," *The Annals of Statistics,* vol. 19, no. 1, pp. 1-67, 3/1 1991, doi: 10.1214/aos/1176347963.

[29] F. Calefato, F. Lanubile, and N. Novielli, "An empirical assessment of best-answer prediction models in technical Q&A sites," *Empirical Softw. Engg.,* vol. 24, no. 2, pp. 854–901, 2019, doi: 10.1007/s10664-018-9642-5.

[30] J. A. Nelder and R. W. M. Wedderburn, "Generalized Linear Models," *Journal of the Royal Statistical Society: Series A (General),* vol. 135, no. 3, pp. 370-384, 1972/05/01 1972, doi: https://doi.org/10.2307/2344614.

[31] W.-Y. Loh, "Classification and regression trees," *WIREs Data Mining and Knowledge Discovery,* vol. 1, no. 1, pp. 14-23, 2011, doi: https://doi.org/10.1002/widm.8.

[32] L. Rokach and O. Maimon, "Decision Trees," in *Data Mining and Knowledge Discovery Handbook*, O. Maimon and L. Rokach Eds. Boston, MA: Springer US, 2005, pp. 165-192.

[33] M. Fernández-Delgado, M. S. Sirsat, E. Cernadas, S. Alawadi, S. Barro, and M. Febrero-Bande, "An extensive experimental survey of regression methods," *Neural Networks,* vol. 111, pp. 11-34, 2019/03/01/ 2019, doi: https://doi.org/10.1016/j.neunet.2018.12.010.

[34] L. I. Kuncheva and C. J. Whitaker, "Measures of Diversity in Classifier Ensembles and Their Relationship with the Ensemble Accuracy," *Machine Learning,* vol. 51, no. 2, pp. 181-207, 2003/05/01 2003, doi: 10.1023/A:1022859003006.

[35] H. Tin Kam, "Random decision forests," in *Proceedings of 3rd International Conference on Document Analysis and Recognition*, 14-16 Aug. 1995 1995, vol. 1, pp. 278-282 vol.1, doi: 10.1109/ICDAR.1995.598994.

[36] J. H. Friedman, "Greedy Function Approximation: A Gradient Boosting Machine," *Annals of Statistics,* vol. 29, pp. 1189--1232, 2000.

[37] T. Chen and C. Guestrin, "XGBoost: A Scalable Tree Boosting System," presented at the Proceedings of the 22nd ACM SIGKDD International Conference on Knowledge Discovery and Data Mining, San Francisco, California, USA, 2016. [Online]. Available: https://doi.org/10.1145/2939672.2939785.

[38] K. Y. Song, I. H. Chang, and H. Pham, "A Testing Coverage Model Based on NHPP Software Reliability Considering the Software Operating Environment and the Sensitivity Analysis," *Mathematics,* vol. 7, no. 5, p. 450, 2019. [Online]. Available: https://www.mdpi.com/2227-7390/7/5/450.

[39] H. Pham, "A New Criterion for Model Selection," *Mathematics,* vol. 7, no. 12, p. 1215, 2019.

[40] S. Luz, F. Haider, S. de la Fuente, D. Fromm, and B. MacWhinney, "Alzheimer's dementia recognition through spontaneous speech: The ADReSS challenge," *Proceedings of the Annual Conference of the International Speech Communication Association, INTERSPEECH,* vol. 2020-Octob, pp. 2172-2176, 2020, doi: 10.21437/Interspeech.2020-2571.

[41] B. A. Kitchenham, E. Mendes, and G. H. Travassos, "Cross versus within-company cost estimation studies: A systematic review," *IEEE Transactions on Software Engineering,* vol. 33, no. 5, pp. 316-329, 2007, doi: 10.1109/TSE.2007.1001.

[42] P. Lenberg, L. G. Wallgren Tengberg, and R. Feldt, "An initial analysis of software engineers' attitudes towards organizational change," *Empirical Software Engineering,* vol. 22, no. 4, pp. 2179-2205, 2017, doi: 10.1007/s10664-016-9482-0.

[43] P. A. Whigham, C. A. Owen, and S. G. MacDonell, "A baseline model for software effort estimation," *ACM Transactions on Software Engineering and Methodology,* vol. 24, no. 3, 2015, doi: 10.1145/2738037.

[44] J. Cohen, *Statistical power analysis for the behavioral sciences*. Routledge, 2013.

[45] M. Kuhn, "Building Predictive Models in R Using the caret Package," *Journal of Statistical Software,* vol. 28, no. 5, pp. 1-26, 2008, doi: 10.18637/jss.v028.i05.

[46] J. W. Cort and M. Kenji, "Advantages of the mean absolute error (MAE) over the root mean square error (RMSE) in assessing average model performance," *Climate Research,* vol. 30, no. 1, pp. 79-82, 2005. [Online]. Available: https://www.int-res.com/abstracts/cr/v30/n1/p79-82.

[47] D. Tien Bui *et al.*, "New Hybrids of ANFIS with Several Optimization Algorithms for Flood Susceptibility Modeling," *Water,* vol. 10, no. 9, p. 1210, 2018. [Online]. Available: https://www.mdpi.com/2073-4441/10/9/1210.

[48] K. O. Maloney, M. Schmid, and D. E. Weller, "Applying additive modelling and gradient boosting to assess the effects of watershed and reach characteristics on riverine assemblages," *Methods in Ecology and Evolution,* vol. 3, no. 1, pp. 116-128, 2012, doi: https://doi.org/10.1111/j.2041-210X.2011.00124.x.

[49] L. Breiman, "Discussion: Multivariate Adaptive Regression Splines," *The Annals of Statistics,* vol. 19, no. 1, pp. 82-91, 1991. [Online]. Available: http://www.jstor.org/stable/2241839.

[50] E. Kartal Koc and H. Bozdogan, "Model selection in multivariate adaptive regression splines (MARS) using information complexity as the fitness function," *Machine Learning,* vol. 101, no. 1, pp. 35-58, 2015/10/01 2015, doi: 10.1007/s10994-014-5440-5.

[51] S. Baltes, R. Kiefer, and S. Diehl, "Attribution required: stack overflow code snippets in GitHub projects," in *Proceedings of the 39th International Conference on Software Engineering Companion*, 2017: IEEE Press, pp. 161-163.

[52] S. M. Nasehi, J. Sillito, F. Maurer, and C. Burns, "What makes a good code example?: A study of programming Q&A in StackOverflow," in *Software Maintenance (ICSM), 2012 28th IEEE International Conference on*, 2012: IEEE, pp. 25-34.

[53] B. Johnson, Y. Song, E. Murphy-Hill, and R. Bowdidge, "Why don't software developers use static analysis tools to find bugs?," in *Software Engineering (ICSE), 2013 35th International Conference on*, 2013: IEEE, pp. 672-681.

[54] N. Ayewah and W. Pugh, "The google findbugs fixit," in *Proceedings of the 19th international symposium on Software testing and analysis*, 2010: ACM, pp. 241-252.

[55] S. M. Lundberg and S.-I. Lee, "A unified approach to interpreting model predictions," presented at the Proceedings of the 31st International Conference on Neural Information Processing Systems, Long Beach, California, USA, 2017.

[56] M. S. Rahman and C. K. Roy, "An Insight into the Reusability of Stack Overflow Code Fragments in Mobile Applications," in *2022 IEEE 16th International Workshop on Software Clones (IWSC)*, 2-2 Oct. 2022 2022, pp. 69-75, doi: 10.1109/IWSC55060.2022.00020.

[57] S. A. Licorish and M. Wagner, "Combining GIN and PMD for code improvements," presented at the Proceedings of the Genetic and Evolutionary Computation Conference Companion, Boston, Massachusetts, 2022. [Online]. Available: https://doi.org/10.1145/3520304.3528772.